\documentstyle[twocolumn,prl,aps]{revtex}
 
\begin{document}
\twocolumn[\hsize\textwidth\columnwidth\hsize\csname
@twocolumnfalse\endcsname
\title{World-Wide Web scaling exponent 
from Simon's 1955 model}
\author{Stefan Bornholdt and Holger Ebel}
\address{Institut f\"ur Theoretische Physik, Universit\"at Kiel,
Leibnizstrasse 15, D-24098 Kiel, Germany
}
\maketitle
\date{today}
\begin{abstract}
Recently, statistical properties of the World-Wide Web have attracted 
considerable attention when self-similar regimes have been observed in 
the scaling of its link structure. 
Here we recall a classical model for general scaling phenomena and 
argue that it offers an explanation for the World-Wide Web's scaling 
exponent when combined with a recent measurement of internet growth. 
\medskip \\ 
\end{abstract}
]

A quantity important for searching the World-Wide Web \cite{butler:2000} 
is the number $k$ of links that point to a 
particular web page. Its probability distribution $P(k)$ exhibits 
power-law scaling 
\cite{barabasi/albert:1999,broder/wiener:2000} 
$P(k) \sim k^{-\gamma}$ 
that is not readily 
explained by standard random graph theory \cite{erdoes/renyi:1960}. 
An elegant model for scaling in copy and growth processes 
was proposed by Simon \cite{simon:1955} in 1955 which 
describes scaling behaviour as observed in 
distributions of word frequencies in texts or population figures 
of cities \cite{zipf:1949}. It models the dynamics of a system 
of elements with associated counters (e.g., words and their frequencies 
in texts, or nodes in a network and their connectivity $k$) where the 
dynamics of the system is based on constant growth via addition 
of new elements (new instances of words) as well as incrementing 
the counters (new occurrences of a word) at a rate proportional 
to their current values.   

Reformulating this to model network growth consider a 
network with $n$ nodes with connectivities $k_i$, $i = 1 \ldots n$, 
forming classes $[k]$ of $f(k)$ nodes with identical connectivity $k$. 
Iterate the following steps: 

(i) With probability $\alpha$ add a new node
and attach a link to it from an arbitrarily chosen node. 

(ii) Else add one link from an arbitrary node to a node 
$j$ of class $[k]$ chosen with probability 
\begin{equation}
P_{\text{new link to class } [k]} \propto k f(k).\label{newlinktoclassk}
\end{equation}

For this stochastic process, Simon finds a stationary solution 
exhibiting power-law scaling with exponent
\begin{equation}
\gamma = 1 + \frac{1}{1- \alpha}\label{scalingexponent}. 
\end{equation}
The only free parameter of the model $\alpha$ reflects the relative  
growth of number of nodes versus number of links. 
In general small values of $\alpha$, therefore,  
predict scaling exponents near $\gamma\approx 2$.  

Let us apply this process to model the evolution of the 
World-Wide Web, identifying nodes with web pages. Data from two 
recent comprehensive Altavista crawls \cite{broder/wiener:2000}
provide an estimate for $\alpha$ in the present internet. 
These two measurements counted 203 million pages and 1466 million 
links in May 1999, and 271 million pages and 2130 million links in 
October 1999. The probability for adding a new web page is estimated 
from the observed increase in counts to $\alpha \simeq 0.10$. 
The subsequent prediction of Simon's model for the exponent of the 
link distribution is $\gamma = 2.1$ comparing well to current 
experimental results $\gamma= 2.1\pm0.1$ 
\protect\cite{barabasi/albert:1999} and $\gamma=2.09$ 
\protect\cite{broder/wiener:2000}. 

To compare with recently proposed models it may be interesting to note that 
the model by Barab\'{a}si and Albert \cite{barabasi/albert:1999} 
can be mapped to the subclass $\alpha= 1/2$ of Simon's model, 
when using the simpler probability for a node being 
connected to another node $i$ with connectivity $k_i$ 
\begin{equation}
P_{\text{new link to i}} \propto k_i \label{prob}. 
\end{equation}
Note that (\ref{prob}) implies (\ref{newlinktoclassk}) 
whereas the reverse is not true. 
Otherwise both models are based on the same two assumptions of 
growth and preferential linking. From this viewpoint, 
it is insightful to reconsider a recent discussion of their model. 
Adamic and Huberman point out that the ``rich-get-richer'' 
behaviour of single nodes imposed by (\ref{prob}) 
correlates age and connectivity of 
nodes \cite{adamic/huberman:2000}. This, however, is disproven 
by data they present. They suggest to (and Barab\'{a}si et al.\ 
in response show how to \cite{barabasi/bianconi:2000}) 
add individual growth rates to each node.  
While this solves the correlation problem, the price to pay is
a large number of free parameters in the extended model. 
A simple solution to this problem has already been provided by Simon: 
Linking is guided by (\ref{newlinktoclassk}) instead of (\ref{prob}),
considering not single nodes but classes of nodes with 
identical connectivities. This allows for different growth 
rates among class members, leaving just one free parameter. 
Above we determine this parameter from experimental data, 
enabling Simon's classical scaling model to estimate the 
connectivity exponent of the World-Wide Web to $\gamma=2.1$. 

\end{document}